# Circular Sphere Decoding for Low Complexity Detection of MIMO Systems with General Two-dimensional Signal Constellations

Hwanchol Jang, Saeid Nooshabadi, *Senior Member,* Kiseon Kim, *Senior Member, IEEE*, and Heung-No Lee[*], *Senior Member, IEEE*

*Abstract*—We propose a low complexity complex valued Sphere Decoding (CV-SD) algorithm, referred to as Circular Sphere Decoding (CSD) which is applicable to multiple-input multiple-output (MIMO) systems with arbitrary two dimensional (2D) constellations. CSD provides a new constraint test. This constraint test is carefully designed so that the element-wise dependency is removed in the metric computation for the test. As a result, the constraint test becomes simple to perform without restriction on its constellation structure. By additionally employing this simple test as a prescreening test, CSD reduces the complexity of the CV-SD search. We show that the complexity reduction is significant while its maximum-likelihood (ML) performance is not compromised. We also provide a powerful tool to estimate the pruning capacity of any particular search tree. Using this tool, we propose the Predict-And-Change strategy which leads to a further complexity reduction in CSD. Extension of the proposed methods to soft output SD is also presented.

## I. Introduction

SPHERE decoding (SD) is a promising multiple-input multiple-output (MIMO) detection strategy because it can achieve the error rate of maximum-likelihood (ML) detection with significantly less complexity compared to the straight-forward ML detector [1], [2]. The standard and most widely used SDs are real valued SDs (RV-SDs) which are only directly applicable to real valued systems. For the usages of RV-SDs, a complex valued system is decomposed into its real and imaginary parts, and it becomes a real valued equivalent system with twice the dimension of its complex valued counterpart. It is worthwhile to note that these standard SDs assume that each data substream is modulated independently. This independency is maintained in the equivalent real valued system when each substream is modulated by a rectangular quadrature amplitude modulation (QAM). Pham *et al.* [3] and Mozos and Garcia [4] state that the application of RV-SD is permissible only for rectangular QAMs because otherwise invalid candidates may arise during the search; it is because the independency between the substream is broken.

It is desirable for an SD to be applicable directly to complex valued systems in the senses of *i*) its flexibility on the choice of signal constellations and *ii*) its efficiency in very large scale integration (VLSI) implementations. *i*): Complex valued SDs (CV-SDs), which are applicable directly to complex valued systems, do not require the decomposition of the systems, and therefore eliminate the limitation of its applicability to independent real and imaginary modulations like a rectangular QAM. Each data substream can be modulated using any two dimensional (2D) constellations. There are many constellations which are desirable to be employed in terms of many aspects of communications performance over rectangular QAMs. The benefits include the reduction in peak-to-average power ratio (PAPR) at each transmit antenna, signal-to-noise ratio (SNR) efficiency, and the increased range of choice in data rate. To list a couple of examples, star QAM reduces the PAPR [5], and near-Gaussian constellations yield a superior shaping gain [6], [7]. Rectangular QAMs depart significantly from these constellations. *ii*): The throughput of VLSI implementation of SD is inversely proportional to the product of the number of visited nodes and the time complexity for each node visit; we assume that one node is visited in each cycle. Burg *et al.* found that the expected number of nodes visited in the SD search is nearly doubled when a complex valued system is decomposed into its equivalent real valued system [8]. For the compensation to this increase, it requires the time complexity for each visit in RV-SD to decrease to half of its CV-SD equivalent. However, the time complexity of RV-SD is almost identical to that of CV-SD [8]. They conclude that CV-SD is the appropriate choice for high throughput VLSI implementations [8].

In CV-SDs, however, the main operation of SD, the pruning test is hard. The test relies on partial Euclidean distance (PED) computations. Here, PED computations are hard, actually the

This work was supported by the National Research Foundation of Korea (NRF) grant funded by the Korean government (MSIP) (NRF-2015R1A2A1A05001826).

This paper was presented in part at the 45th Asilomar Conference on Signals, Systems and Computers, Pacific Grove, CA, 6-9, Nov. 2011

Hwanchol Jang was pursuing Ph.D. at Gwangju Institute of Science and Technology (GIST) while completing this work, and is currently with Advanced Photonics Research Institute, GIST, Gwangju, 500-712 Korea, e-mail: hcjang@gist.ac.kr. Saeid Nooshabadi is with the Department of Electrical and Computer Engineering and the Department of Computer Science, the Michigan Technological University, Houghton, MI, 49931, e-mail:saeid@mtu.edu. Kiseon Kim and Heung-No Lee are with the School of Information and Communications, GIST, e-mail: {kskim, heungno}@gist.ac.kr. The asterisk * indicates the corresponding author.

most expensive operations in SD, and thus, the CV-SD has high complexity. In previous CV-SDs, it is considered to restrict the applicable constellation to *i*) those whose elements are aligned in concentric rings of different radii [3], [4], [8], [9] and *ii*) those whose elements are aligned with several vertical or horizontal lines [10] for complexity reduction. Nevertheless, the general CV-SD without such restriction on its constellation is still remained to be computationally expensive.

It is our goal in this paper, therefore, to develop a low complexity CV-SD for general 2D constellations. We aim to do this while guaranteeing the ML performance. To this end, we take a new approach which takes advantage of a simple necessary condition, rather than an equivalent one, of the original pruning constraint, the sphere constraint (SC).

The main contributions of this paper are summarized as follows:
- We derive a necessary condition with which the metric for the constraint test becomes the magnitude of a scalar, not the Euclidean distance of a vector, thus the constraint test becomes very simple.
- We use the simpler metric and devise a novel complexity reduced CV-SD algorithm. The proposed constraint and the proposed CV-SD algorithm employing the constraint is referred to as the circular constraint (CC) and circular sphere decoding (CSD), respectively. CSD employs a two-step constraint test. In the prescreening step, those constellation points which are not promising are eliminated by the simple CC tests. The pruning by SC tests, in the second step, requiring expensive PED calculations is performed only for those candidates which have survived the CC tests. Thus, many expensive PED computations are avoided. Significant savings in the complexity of the CV-SD can be made with CSD without sacrificing the ML performance.
- We also propose a novel Predict-And-Change (PAC) strategy which further utilizes CC and reorganizes the tree so that tree pruning close to the root is increased. This leads to a substantial complexity reduction in CSD.
- We provide an extension of CSD for soft output SD for coded MIMO systems.

The rest of this paper is organized as follows. In Section II, the system model is presented. In Section III, the underlying principle in SD, and the difference between RV-SD and CV-SD are studied. In Section IV and V, the proposed CSD algorithm and the proposed PAC strategy are developed. In Section VI, an extension of CSD to list SD is provided. In Section VII, the complexity analysis for the proposed CSD is given. In Section VIII, we discuss the system simulation results. Section IX concludes the paper.

## II. SYSTEM MODEL AND NOTATION

We consider a complex valued baseband MIMO channel model with $m$ receive and $n$ transmit antennas ($m \geq n$). Consider the system model,

$$\mathbf{r} = \mathbf{H}\overline{\mathbf{s}} + \mathbf{v}, \qquad (1)$$

where $\mathbf{r} \in \mathbb{C}^m$ denotes the received signal, $\mathbf{H} \in \mathbb{C}^{m \times n}$ denotes the $m \times n$ block Rayleigh fading channel matrix whose entries are independent and identically distributed (i.i.d.) circularly-symmetric complex Gaussian (CSCG) random variables $\mathcal{CN}(0,1)$, $\overline{\mathbf{s}} \in \mathcal{O}^n \subset \mathbb{C}^n$ is the transmitted symbol vector where $\mathcal{O} := \{o_1, o_2, \cdots, o_L\}$ can be any discrete 2D constellation set with size $L$; $o_i \in \mathbb{C}$ for $1 \leq i \leq L$. The components of $\overline{\mathbf{s}}$ are i.i.d. and take the values uniformly from $\mathcal{O}$, $\overline{s}_k \in \mathcal{O}$, and they are scaled to have $\mathrm{E}|\overline{s}_k|^2 = 1$ for $\forall 1 \leq k \leq n$. The $m \times 1$ vector $\mathbf{v} \in \mathbb{C}^m$ is the additive noise whose entries are i.i.d. CSCG random variables $\mathcal{CN}(0,\sigma^2)$. The channel $\mathbf{H}$ is assumed to be known at the receiver.

Variables that denote vectors and matrices are set, respectively, lowercase and uppercase boldface. $\mathbf{H}_{k,\_}$ and $\mathbf{H}_{\_,j}$ denotes the $k^{\mathrm{th}}$ row and the $j^{\mathrm{th}}$ column of matrix $\mathbf{H}$, respectively. $\mathbf{H}^*$ and $\mathbf{H}^\dagger$ denote the conjugate transpose and the pseudo-inverse of $\mathbf{H}$, respectively. An individual component of a matrix or a vector is identified by the subscript. For example, $H_{i,j}$ and $s_k$ are the (*i*,*j*) component of $\mathbf{H}$ and, the $k^{\mathrm{th}}$ component of $\mathbf{s}$, respectively. $\mathbf{s}_{k:N}$ is the vector taking the last $n-k+1$ components of $\mathbf{s}$. $|\mathcal{O}|$ denotes the cardinality of the set $\mathcal{O}$. $|a|$ denotes the magnitude of $a \in \mathbb{C}$. $\|\mathbf{s}\|$ denotes the $2^{\mathrm{nd}}$ norm of vector $\mathbf{s}$. $\mathbb{R}$ and $\mathbb{C}$ denote the real and the complex domains respectively. $\Re(s_k)$ and $\Im(s_k)$ denote the real and the imaginary parts of $s_k$ respectively. $(a)_+$ is $a$ if $a>0$, and 0 otherwise.

## III. COMPARISON BETWEEN RV-SD AND CV-SD

In this section, we describe the SD principle and the complexity problem in CV-SD.

### A. SD principle

The standard procedure of SD is *i*) to identify all the candidates $\mathbf{s}$ that satisfy the sphere constraint (SC), and *ii*) to choose the candidate with the minimum distance to the received signal $\mathbf{r}$ as the solution. The SC is expressed by

$$d(\mathbf{s}) := \|\mathbf{r} - \mathbf{H}\mathbf{s}\|^2 \leq C, \qquad (2)$$

where $\mathbf{s} \in \mathcal{O}^n$. But, this is not efficient for implementation. For a better implementation, the SC can be expressed as

$$\begin{aligned} d_k(\mathbf{s}_{k:n}) &:= \sum_{i=k}^{n} \left| y_i - \mathbf{R}_{i,i:n}\mathbf{s}_{i:n} \right|^2 \\ &= d_{k+1}(\mathbf{s}_{k+1:n}) + \left| b_{k+1}(\mathbf{s}_{k+1:n}) - R_{k,k}s_k \right|^2 \leq C, \end{aligned} \qquad (3)$$

for $1 \leq k \leq n$ where $\mathbf{R}$ is the upper triangular matrix from the QR decomposition of $\mathbf{H}$, $\mathbf{H}=\mathbf{QR}$, $\mathbf{y} := \mathbf{Q}^*\mathbf{r}$, $d_{n+1} = 0$, and $b_{k+1}(\mathbf{s}_{k+1:n}) := y_k - \mathbf{R}_{k,k+1:n}\mathbf{s}_{k+1:n}$. Here, $d_k(\mathbf{s}_{k:n})$ is referred to as partial Euclidean distance (PED) and it depends on the partial vector $\mathbf{s}_{k:n}$, the last $n-k+1$ components of $\mathbf{s}$, which is associated with the nodes at the $k^{th}$ level of the tree. As PED monotonically increases as $k$ decreases, PEDs for the remaining levels of $k$ do not need to be computed once PED of a level is found to violate the SC in (3). This gives computational savings to SD. As a result, SDs provide complexity reductions for ML solution search.

*Definition 1: (element-wise independent metric)* We call a metric is *element-wise independent* if the metric at a level of tree, say the $k^{th}$ level, depends only on the corresponding element $s_k$ in $\mathbf{s}$ but not on any other $s_i$ where $i \in \{1,2,\cdots,n\}$ and $i \neq k$. That is, among all the elements of $\mathbf{s}$, $s_k$ is the only argument of the metric.

Here, we should note that the SC tests in (3) are still hard as the PEDs are element-wise dependent. Assume that the value of $s_i$ for any $1 \leq i \leq n$ changes, the PED of the $k^{th}$ level of the tree for level $k \leq i$ need to be calculated again. This incurs the number of required PED computations and the cost for each PED computation increased exponentially as $k$ decreases. In the past, a simplified SC where explicit PED computations are not required was employed for a lower complexity algorithm. Further simplification on SC was obtained, as noted in the introduction, by employing rectangular QAMs and exploiting the characteristics that exists in the constellations, e. g., the independence between the real and the imaginary components.

*B. The difference between RV-SD and CV-SD*

In real valued systems where the components of $\mathbf{r}$, $\mathbf{H}$, $\bar{\mathbf{s}}$, $\mathbf{s}$, and $\mathbf{v}$ are real valued, the SC in (3) can be simplified by the so called admissible interval (AI) which is expressed by the lower limit $s_k^l$ and the upper limit $s_k^u$ as follows

$$s_k \in \left[ s_k^l, s_k^u \right], \quad (4)$$

where $s_k^l = \frac{b_{k+1}(\mathbf{s}_{k+1:n})}{R_{k,k}} - \frac{\sqrt{C-d_{k+1}(\mathbf{s}_{k+1:n})}}{R_{k,k}}$ and $s_k^u = \frac{b_{k+1}(\mathbf{s}_{k+1:n})}{R_{k,k}} + \frac{\sqrt{C-d_{k+1}(\mathbf{s}_{k+1:n})}}{R_{k,k}}$. The AI is still element-wise dependent since $b_{k+1}(\mathbf{s}_{k+1:n})$ and $d_{k+1}(\mathbf{s}_{k+1:n})$, for the AI at the $k^{th}$ level of the tree, are functions of $\mathbf{s}_{k+1:n}$, other than $s_k$. But it can be identified by calculating only the two values, $s_k^l$ and $s_k^u$. The SC of (4) is used in RV-SD instead of the SC of (3) since it is simpler. The constellation points $s_k$ which satisfy the SC can be identified without explicit PED computations $d_k(\mathbf{s}_{k:n})$ of candidates $\mathbf{s}_{k:n}$ for the current level $k$ of the tree. PED computations are needed only for the nodes in the AI in RV-SD. They are not for pruning itself but for setting the AI, $s_{k-1}^l$ and $s_{k-1}^u$, for the next level of the tree. However, note that, the AI simplification of the SC applies only to rectangular QAMs. References [1], [2] should be consulted for a more conceptual description of RV-SD.

In this work, the aim is for a CV-SD for general 2D constellations. This makes it difficult to replace the expensive operations for the SC in (4) with ones that are much cheaper. Hence, every single pruning in the CV-SD is done by the explicit expensive PED computations for the SC test in (3) [8]. This results in high complexity in the CV-SD.

Let us see the difference on the number of PED computations of the CV-SD and the RV-SD, and the numbers of the floating point operations (FLOPs) of them. Consider a tree which is used for the SD search. The tree expands with the factor $L$ as the level of the tree, $k$, goes down, starting from $k = n$ to $k = 1$. The number of nodes at the $k^{th}$ level of the tree is $L^{n-k+1}$ and each node represents a candidate for the partial vector $\mathbf{s}_{k:n}$. Let $N_{k+1}^{sc}$ be the number of nodes that satisfy the SC at $(k+1)^{th}$ level of the tree. In the CV-SD, PEDs are calculated for all the children nodes of the surviving nodes. That is $L \cdot N_{k+1}^{sc}$ PED computations for the $k^{th}$ level of the tree. Thus, the number of PED computations is expanded by $L$ in CV-SD. We name this *L-expansion property* of the CV-SD. This property is trivial but it will be easier for us to recall this property later on. In the RV-SD, only the $N_k^{sc}$ nodes which are inside the AI are required for PED computations at the $k^{th}$ level of the tree no matter how large $N_{k+1}^{sc}$ is. Surely, $L \cdot N_{k+1}^{sc} \geq N_k^{sc}$. The respective FLOPs of the PED computations for CV-SD and RV-SD are $\left(6(n-k)+8L\right)N_{k+1}^{sc}$ and $\left(2(n-k)+4\right)N_k^{sc}$. [1] Actually, the direct comparisons of the numbers of the PED computations and those of the FLOPs are not fair for the system models for RV-SD and for CV-SD are different. But, at least, we can see the inefficiency of the CV-SD for it includes $L \cdot N_{k+1}^{sc}$ factor rather than $N_k^{sc}$.

For low complexity CV-SDs, Hochwald and Brink [9], Burg *et al.* [8], Pham *et al.* [3], and Mozos and Garcia [4] consider restricting the applicable signal constellations only to those whose elements are aligned in several concentric rings with different sizes, rather than general 2D constellations, so that they can exploit the constellation structure. An AI is obtained for these complex valued constellations. There, the interval is not for the value of $s_k$ itself but for its phase $\angle s_k$. Obviously the applicability of the method is limited to only those constellations with the specific shape. In addition, it requires costly trigonometric function and other computations. In this paper, we do not consider this approach.

*C. Schnorr-Euchner enumeration in RV-SD and CV-SD*

In Schnorr-Euchner (SE) enumeration, the children nodes of a parent node are visited in the ascending order of their PEDs during the search. Therefore, once a preceding child node is found to violate the SC, it is definite that the remaining siblings also violate the SC and they do not need to be searched. This provides a considerable complexity reduction to SD.

---

[1] Computation of $b_{k+1}(\mathbf{s}_{k+1:n})$ requires $(n-k)$ complex multiplications and $(n-k)$ complex additions, totaling $6(n-k)$ FLOPs (one complex multiplication and one complex addition are equivalent to four FLOPs and two FLOPs, respectively). The remaining computation for the PED computation in (3) requires 8 FLOPs. The computation of $b_{k+1}(\mathbf{s}_{k+1:n})$ is required only once for it is common for the candidates as it does not depends on $s_k$. For RV-SD, computation of $b_{k+1}(\mathbf{s}_{k+1:n})$ requires $2(n-k)$ FLOPs and the remaining computation requires 4 FLOPs. Note that one more FLOP is required in case of a SC test for a comparison with $C$; consider this in Sec. IV-A.

In RV-SD, the sorting can be performed without explicit PED computations of the siblings. The first sibling $s_k$ is determined by slicing $\frac{b_{k+1}(s_{k+1:n})}{R_{k,k}}$ to the closest constellation point. The sequence of the remaining siblings is then determined by the zigzag ordering of the neighboring constellation points of the first sibling [1]. Thus, the SE enumeration in RV-SD can be done very efficiently.

To the best knowledge of the authors, there is no efficient SE enumeration scheme for the CV-SD which is applicable to arbitrary 2D constellations and achieves the exact ML performance; we consider this in this paper. Thus, the SE enumeration for the CV-SDs requires explicit PED computations for all the siblings. Note that there are several efficient SE or SE-like enumeration schemes [3], [4], [8], [10], [11]. But, they are for the CV-SDs which *i*) limit their constellations to certain kinds and exploit the special structures of them, and/or *ii*) compromise the exact ML performance. *i*): Pham *et al.* [3], Mozos and Garcia [4], and Burg *et al.* [8] apply SE enumerations on the constellations whose elements are aligned in several concentric rings with different sizes. The SE enumeration by Hess *et al.* [10] partitions the constellation into subsets consisting of rows or columns and reduces the enumeration overhead. But, this is efficient only for the constellations most of whose elements are aligned with several vertical or horizontal lines. *ii*): The SE-like method by Wenk *et al.* [11] uses approximations, such as the *l*-1 norm or the so called *l*-infinity norm, of the PED for efficient enumeration. In this method, the remaining siblings are pruned once a preceded sibling node is found to violate the SC as they are in other SE enumerations. This may prune the node from which the ML solution originates; the method uses an approximation, not the PED itself.

## IV. PROPOSED CIRCULAR SPHERE DECODING

The applicability to general 2D constellations is an important benefit of the general CV-SD. But, this generality gives an inherent problem in the CV-SD that no structure of a specific constellation can be exploited for a simplification of the SC. This makes it difficult to have a simpler but equivalent constraint to the SC in (3).

In this section, we simplify the SC of (3) by resorting to one of its necessary conditions, rather than to any specific structure in constellations. We refer to this necessary condition as the circular constraint (CC). The CC may not prune as many nodes as the SC does since the CC is a necessary condition. In order not to lose any pruning efficiency, we propose that SC tests are executed for those nodes which are not eliminated by the preceded CC tests. We call this CV-SD that employs CC tests circular sphere decoding (CSD). CSD prunes the nodes the same amount as the baseline CV-SD does but with a smaller number of hard SC tests.

### A. Circular constraint (CC)

SC tests in (3) are hard due to the element-wise dependence of the PEDs. Now, we aim to find a new constraint where the element-wise dependence of the metric is removed. We start from the SC in (3). The element-wise dependency is removed by eliminating the matrix $\mathbf{R}$ inside the norm operator. The derivation is given by

$$\begin{aligned} C &\geq d_k(\mathbf{s}_{k:n}) \\ &= \left\| \mathbf{R}_{k:n,k:n}(\mathbf{x}_{k:n} - \mathbf{s}_{k:n}) \right\|^2 \\ &= \frac{\left\| \mathbf{H}^\dagger_{k,-} \right\|^2}{\left\| \mathbf{H}^\dagger_{k,-} \right\|^2} \left\| \mathbf{R}_{k:n,k:n}(\mathbf{x}_{k:n} - \mathbf{s}_{k:n}) \right\|^2 \\ &\underset{(a)}{\geq} \frac{\left| \mathbf{R}^\dagger_{k,k:n} \mathbf{R}_{k:n,k:n}(\mathbf{x}_{k:n} - \mathbf{s}_{k:n}) \right|^2}{\left\| \mathbf{H}^\dagger_{k,-} \right\|^2} \\ &\underset{(b)}{=} \frac{\left| x_k - s_k \right|^2}{\left\| \mathbf{H}^\dagger_{k,-} \right\|^2}, \end{aligned} \quad (5)$$

where $\mathbf{x} := \mathbf{H}^\dagger \mathbf{r}$ with $\mathbf{H}^\dagger := (\mathbf{H}^*\mathbf{H})^{-1} \mathbf{H}^*$, (a) is from $\left\| \mathbf{H}^\dagger_{k,-} \right\| = \left\| \mathbf{R}^\dagger_{k,k:n} \right\|$ and the Cauchy-Schwarz inequality, and (b) is from the fact that $\mathbf{R}^\dagger \mathbf{R}(\mathbf{x}-\mathbf{s}) = \mathbf{x}-\mathbf{s}$ with the assumption that $\mathbf{H} \in \mathbb{C}^{m \times n}$ has the full rank with $m \geq n$. For a $\mathbf{H}$ with rank deficiency, there are contributions in the metric derived in (5) from the elements of $\mathbf{x}-\mathbf{s}$ other than the $k^{\text{th}}$ element of it. But, they are insignificant unless $\mathbf{H}$ has serious rank deficiency.

Now, we have a new constraint, CC,

$$\Delta_k(s_k) \leq C \cdot \delta_k^2, \ k = 1, 2, \cdots, n, \quad (6)$$

where $\Delta_k(s_k) := \left| x_k - s_k \right|^2$ and $\delta_k^2 := \left\| \mathbf{H}^\dagger_{k,-} \right\|^2$. We name the metric $\Delta_k(s_k)$ as circular metric (C-metric). Here, the value of $\delta_k^2$ can be computed before the SD search begins, and remains unchanged as long as $\mathbf{H}$ does not change., thus it can be used while $\mathbf{H}$ stays the same. Note that CC is a necessary condition for a candidate $\mathbf{s}$ to satisfy the SC. It is because the metric derived in (5) is smaller than or equal to the PED for SC in (3).

The C-metric $\Delta_k(s_k)$ is element-wise independent since it depends only on $s_k$ (Def. 1). This element-wise independence gives CC two beneficial features in terms of complexity. First, the required number of CC tests is fixed only to $L$ for each level of the tree. It is because the C-metric does not depend on the elements of the parent node, and hence the C-metrics for the children nodes originated from one parent node are the same with those from any other parent nodes. Note that the number of required SC tests at the $k^{\text{th}}$ level of the tree is $L \cdot N^{sc}_{k+1}$ (Sec. III-B); this ranges from $L$ to $L^{n-k+1}$. Second, each CC test is simple. A CC test requires only six FLOPs while a SC test requires $6(n-k)+9$ FLOPs for the first sibling at the $k^{\text{th}}$ level of the tree and 9 FLOPs for the remaining siblings. [1] Thanks to these two features, the CC tests at the $k^{\text{th}}$ level of the tree require only $6L$ FLOPs while the SC tests require $(6(n-k)+9L)N^{sc}_{k+1}$ FLOPs.

Note that there are some lower bounds used in other SDs [12],

[13]. Stojnic *et al.* provide several lower bounds on the remaining path metric using ideas from $H^\infty$ estimation theory and some of its special cases [12]. Barik and Vikalo obtain a lower bound on the metric by relaxing the metric minimization problem [13]. But, it is difficult to utilize them for prescreening on the SC tests because *i*) they are not lower bounds on the current PED but on the remaining path metric and *ii*) they are computationally expensive due to their element-wise dependent metric computations; they require $b_{k+1}(\mathbf{s}_{k+1:n})$.

### B. Circular sphere decoding

In CSD, we utilize the simple CC for prescreening on the SC tests, thus reduce the computational complexity of the search. The strategy in CSD is *i*) to eliminate as many nodes as possible for a given level of the tree using the CC, and then *ii*) to perform SC tests only for the surviving nodes.

Consider Fig. 1 where we illustrate the CSD operations by providing a geometrical presentation of CC and SC. The CC is represented in the **s** space by *n* separate circles, one for each element of **s**. The SC is represented by the sphere in **Hs** space. In the prescreening step of CSD, the constellation points which are not inside each separate circle are excluded from the search. In the pruning step, SC tests are performed only for the prescreened candidates and vector points which are inside the sphere are identified. In CSD, many non-promising candidates are eliminated in the prescreening test even before their SC tests are performed. As it is shown in Fig. 1, only a portion of the points are pruned by the SC tests in CSD (Fig. 1 (b)), while whole pruning operations in the CV-SD are solely through the SC tests. Thus, the employment of the CC test in CSD reduces the complexity of the CV-SD. As a result, CSD outperforms the CV-SD in terms of complexity (Section VIII). Note that the ML performance in CSD is not compromised since the CC test does not eliminate the ML solution; the CC is necessary for a candidate for the satisfaction of the SC.

The proposed CSD is a quite general MIMO detection framework that can be used together with other advantageous complexity reduction techniques or can be employed in other kinds of MIMO detection problems for further benefits. CSD can be employed with statistical pruning techniques such as that in [14] for additional complexity reduction. This can be done just by replacing *C* by one in [14]. Note also that CSD can be used to benefit those MIMO schemes employing space-time block codes in [15] and references therein. They have additional zero elements in the triangular matrix **R** of their equivalent system models. The zero elements reduce the element-wise dependency in their PED computations. However, the PEDs are still not fully element-wise independent. CSD, with element-wise independent metric, can thus be applied to those additionally structured MIMO systems and provide further complexity reduction for ML detection.

The entire algorithm of the proposed CSD is given in Table I.

### C. Circular enumeration

SE enumeration has a distinct benefit in reducing the complexity of the search. But, the SE enumeration for general CV-SDs is not suitable for CSD due to its heavy overhead. Fortunately, there are C-metrics available in CSD. C-metric of a node is a kind of a lower bound on the PED of the node (Eq. (5)). Thus, it can be used as a surrogate PED for the purpose of efficient enumeration.

We sort the constellation points in non-decreasing order with respect to their C-metrics as follows,

$$s_k^1, s_k^2, \cdots, s_k^L \text{ s. t. } \Delta_k(s_k^1) \leq \Delta_k(s_k^2) \leq \cdots \leq \Delta_k(s_k^L). \quad (7)$$

This provides the benefit of making more nodes be pruned by the efficient CC tests. The CC becomes stricter whenever the search proceeds and reaches a leaf in the tree, where the radius is reduced to the metric for the leaf. Now that the siblings are

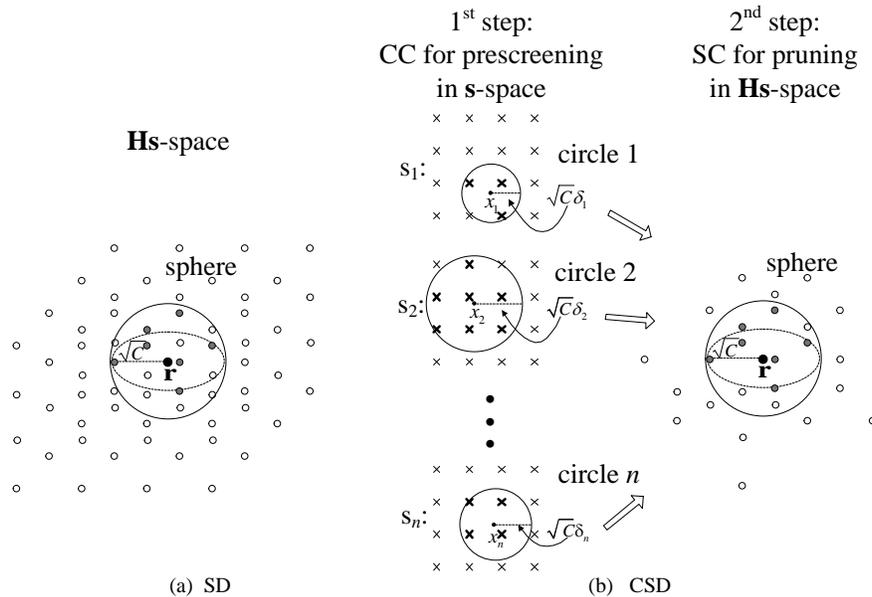

Fig. 1. Geometry of SD and the proposed CSD with 16 QAM constellation (gray-colored points represent those points which are inside the sphere). (a) SD. (b) CSD.

TABLE I
**The CSD Algorithm**

**Definitions**: $\mathcal{O} = \{o_1, o_2, ..., o_L\}$ (the constellation set), $\boldsymbol{\delta}^2 = [\delta_1^2, \delta_2^2, ..., \delta_n^2]^T$, $k$ (the current level), $C$ (squared radius), $\mathbf{s} = [s_1, s_2, ..., s_n]^T$ (the currently visited node), $d_k$ (the PED for the current level), $\hat{\mathbf{s}}$ (ML solution).

**Inputs**: $\mathcal{O}$, $\boldsymbol{\delta}^2$, $C_0$, $\mathbf{r}$, $\mathbf{x}$, and $\mathbf{R}$.

**Outputs**: $\hat{\mathbf{s}}$.

**Step 0**: (C-metric) Compute C-metrics $\Delta_k(s_k)$ for $\forall s_k \in \mathcal{O}$ and $\forall k$.

**Step 1**: (Initialization)
    set $C \leftarrow C_0$, $k \leftarrow n$, $\text{found} \leftarrow 0$.

**Step 2**: (Initialization of the node for a visit)
    **for** $k = 1:n$
        set $I_k = 1$ and $s_k = o_1$.
    **end**
    Go to step 4

**Step 3**: (Next node)
    **if** $I_k \neq L$, set $I_k = I_k + 1$ and $s_k = o_{I_k}$, and go to step 4
    **else** go to step 8

**Step 4**: (CC test *)
    **if** $\Delta_k(s_k) \leq C \cdot \delta_k^2$, go to step 5
    **else** go to step 3

**Step 5**: (PED) Compute PED $d_k(\mathbf{s}_{k:n})$.

**Step 6**: (SC test)
    **if** $d_k(\mathbf{s}_{k:n}) < C$ go to step 7
    **else** go to step 3

**Step 7**: (Forward)
    **if** $k = 1$, $C \leftarrow d_k$ (radius updates), $\hat{\mathbf{s}} \leftarrow \mathbf{s}$, $\text{found} \leftarrow 1$, go to step 3.
    **else** $k \leftarrow k - 1$, go to step 4

**Step 8**: (Backward)
    **if** $k = n$ (root node)
        **if** $\text{found} = 0$, $C_0 \leftarrow C_0 + \Delta C_0$ (radius increase), go to step 2
        **else** exit
    **else**
        **for** $k' = 1:k$
            set $I_{k'} = 1$ and $s_{k'} = o_1$.
        **end**
        $k \leftarrow k + 1$, go to step 3

\* $C \cdot \delta_k^2$ is not computed for every CC test. It is computed only when the radius $C$ is updated in Step 7.

sorted by the C-metric, stricter CC tests are performed for those siblings with larger C-metrics which have higher chances of being pruned by the CC. This results more candidates to be eliminated before the SC tests. It will be later shown in Sec. VIII that CSD with the circular enumeration (C-CSD) provides considerable complexity reduction to CSD; it also outperforms the SE-SD.

## V. PROPOSED PREDICT-AND-CHANGE STRATEGY

Pruning of a single node at a level of a tree amounts to pruning of its whole underlying sub-tree whose size in the number of nodes is exponential to the number of levels left to reach to the leaf level. It is thus efficient in all SDs, including CSD, to prune nodes at higher levels of the tree. Aiming to increase pruning at higher levels of the tree, we take the approach which *i*) predicts the pruning potential of each symbol $s_i$ of $\mathbf{s}$ for $1 \leq i \leq n$, and *ii*) reorganizes the tree so that the symbols with larger pruning potentials to be placed at higher levels of the tree; the CSD search is performed on this tree. In the following subsection, we develop the idea of pruning potential for each symbol and provide a method to calculate it.

### A. Symbol pruning potential

The pruning potential we propose to use here is an upper bound on the number of constellation points that can be pruned by the CC test. The largest number of constellation points that can be pruned by the CC test is calculated by using the strictest CC test which prunes as many constellation points as possible but without pruning the ML solution. Of course, this requires the identification of the ML solution or $d(\mathbf{s}_{\text{ML}})$ which are not available until the CSD search is completed. That is why we instead obtain an upper bound. This can be calculated by using the concept of the minimum circles. Note that so far there is no computationally feasible way to predict a non-trivial upper bound on the number of nodes that can be pruned by the SC test prior to a SD search.

*Definition 2: (the minimum circles)* The minimum circles (MCs) are the $n$ smallest circles *i*) which are centered at $x_1, x_2, \cdots, x_n$, respectively, *ii*) the proportion of whose radii is $\delta_1 : \delta_2 : \cdots : \delta_n$, and *iii*) each of which contains at least a single constellation point inside it. The MCs are the geometric view of the strictest CC test which is satisfied by at least one of $\mathbf{s} \in \mathcal{O}^n$.

Let us denote the set of constellation points inside the $i^{\text{th}}$ circle of the MCs by $\mathcal{O}_i^{\text{MC}}$. $\mathcal{O}_i^{\text{MC}}$ can be identified by using the following proposition.

*Proposition 3:* A constellation point $s_i \in \mathcal{O}$ belongs to $\mathcal{O}_i^{\text{MC}}$ if and only if
$$\Delta_i(s_i) \leq C_{\min} \cdot \delta_i^2, \tag{8}$$
where $C_{\min} := \max_i \left\{ \delta_i^{-2} \min_{s_i \in \mathcal{O}} \Delta_i(s_i) \right\}$.

*Proof:* The minimum radius for each circle of the MCs is $\delta_i^{-2} \min_{s_i \in \mathcal{O}} \Delta_i(s_i)$ for $1 \leq i \leq n$. The maximum radius is selected to guarantee that all the circles in MCs contain at least one constellation point. □

The CC test corresponding to the MCs is stricter than or equal to the strictest CC test which the ML solution satisfies. Note that the ML solution is not guaranteed to be included in the vector points $\mathbf{s}$ which are constituted by the constellation points inside the MCs. Thus, we can obtain the pruning potential $P_i$, an upper bound on the number of constellation points that can be pruned by the CC test, for each symbol at $1 \leq i \leq n$, as follows
$$P_i := L - \left| \mathcal{O}_i^{\text{MC}} \right|$$
$$= \left| \{ s_i | \Delta_i(s_i) > C_{\min} \delta_i^2, \ s_i \in \mathcal{O} \} \right|. \tag{9}$$

## B. Predict-And-Change (PAC) strategy

The pruning potential $P_i$ for each symbol $s_i$ for $i = 1, 2, ..., n$ in **s** can be obtained using (9) once the C-metrics are computed for a received signal **r**. Using the pruning potentials, the symbols in **s** and the columns of **H** are reordered. We propose that the $n$ symbols in **s** are placed from the root ($k = n$) to the leaf ($k = 1$) of the tree in non-increasing order with respect to their pruning potentials. The corresponding $n$ columns in **H** are also reordered accordingly. The reordered **s**′ and **H**′ are constituted as follows,

$$\mathbf{s}' = [s_{i_1}, s_{i_2}, \cdots, s_{i_n}]^T \text{ and } \mathbf{H}' = [\mathbf{H}_{\_,i_1}, \mathbf{H}_{\_,i_2}, \cdots, \mathbf{H}_{\_,i_n}]$$
$$\text{s. t. } P_{i_1} \leq P_{i_2} \leq \cdots \leq P_{i_n}. \quad (10)$$

Then, the reorganized search tree is created by doing the QR decomposition on **H**′. The CSD search is performed on the reorganized tree; the reorganization is done right before Step 1 in TABLE I.

Different from the SD ordering schemes which consider only **H** [1], [2], PAC exploits the information of the received signal **r**. Note that this becomes possible in PAC because it is based on the pruning potentials which are available prior to the formation of a tree. It is shown in Fig. 2 that the complexity reduction in CSD which employs the PAC strategy (PAC-CCSD) is substantial compared to that with the conventional ordering (PINV-CCSD); this conventional ordering places the symbol with a smaller inverse channel norm $\|\mathbf{H}^\dagger_{i,\_}\|^2$ at a higher level of the tree.

## C. A modified PAC

QR decomposition in SD needs to be performed whenever the channel **H** changes. For PAC strategy, it requires $n$ factorial QR decompositions per channel change since there exist $n$ factorial different reordered channel matrices **H**′. This may not be any problem in terms of computational overhead when the channel changes slowly (quasi-static channel), but otherwise, it may become problematic.

We here provide a variation of PAC and reduce the computational overhead of QR decompositions. This reduces the overhead per channel change significantly while the benefit of PAC per channel use is still kept large. For the variation, we consider a subset of **H**′, only $n$ different reordered channel matrices **H**′ out of the total $n$ factorial of them. The reorganization is modified as follows,

$$\mathbf{s}' = [s_{i_1}, s_{i_2}, \cdots, s_{i_n}]^T \text{ and } \mathbf{H}' = [\mathbf{H}_{\_,i_1}, \mathbf{H}_{\_,i_2}, \cdots, \mathbf{H}_{\_,i_n}]$$
$$\text{s. t. } i_n = \arg\max_{i_k}\{P_{i_k}\} \text{ and } \delta^2_{i_1} \geq \delta^2_{i_2} \geq \cdots \geq \delta^2_{i_{n-1}}. \quad (11)$$

The root of the tree, $k = n$, is placed by the symbol with the largest pruning potential. The other levels of the tree from $k = n-1$ to $k = 1$ are placed by the remaining symbols in non-decreasing order with respect to their $\delta^2_i$.

Although the pruning potentials are utilized only for the symbol to be placed at the root and $\delta^2_i$ is utilized instead for the other levels of the tree, the benefit of PAC is reduced only slightly (Sec. VIII). The intuition for this is that *i*) we use the pruning potential at the level with the best pruning efficiency and *ii*) for other levels, we still consider pruning potentials but in the average sense since a smaller $\delta^2_i$ indicates a larger $P_i$ in the case where the information on $\Delta_i(s_i)$ and $C_{\min}$ is not available (refer to (9)).

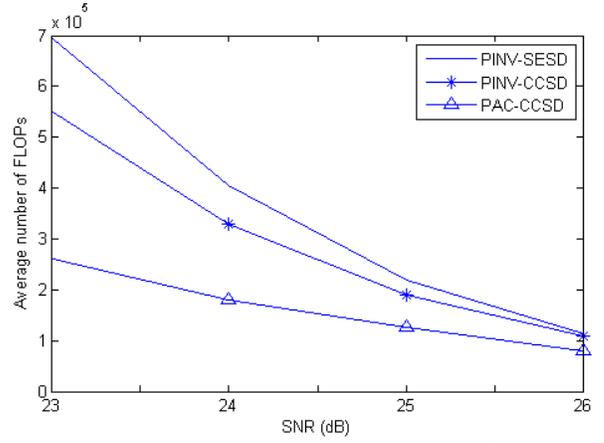

Fig. 2. Ordering benefits: complexity of PAC-CCSD, PINV-CCSD, and PINV-SESD for 10×10 MIMO systems with 64 QAM.

It is now possible with these $n$ QR decompositions to give a comparable complexity reduction to that of the $n$ factorial QR decompositions. In addition, we provide in the following subsection an efficient way of performing the $n$ QR decompositions and the computational overhead is reduced to be less than that of two QR decompositions. With the modification and the efficient QR decompositions, we can say that the problem of PAC with the computational overhead is resolved.

For a clear description on the benefit of the modified PAC, we provide a comparison of complexities for the conventional PINV ordering based SE-SD (PINV-SESD), the PAC based C-CSD (PAC-CCSD), and the modified PAC based C-CSD (PAC*-CCSD) in TABLE II. They are for $n=m=10$ with (8,24,32) star 64 QAM. We split the complexities into those for channel rate processing and those for symbol rate processing. This is because the overhead is incurred in the channel rate preprocessing stage and the benefit is obtained in the symbol rate detection stage. This separately shows the gains and the losses of the PAC-CCSD and the PAC*-CCSD compared to PINV-SESD. We also provide the number of channel uses, $N_{\text{CH}}$, for PAC-CCSD and PAC*-CCSD to be net beneficial. It is found that PAC-CCSD is the most advantageous in symbol rate detection. However, the net benefit to PINV-SESD is obtained in a very restricted condition where $N_{\text{CH}} > 106619$. It is seen

TABLE II

| Ordering | Channel-rate preprocessing | Per-Symbol detection | $N_{\text{CH}}$ |
|---|---|---|---|
| PINV | $C_{\text{QR}} \sim 5.33 \times 10^3$ | $C_{\text{PINV-SESD}} \sim 4.33 \times 10^5$ | |
| PAC | $n!\ C_{\text{QR}} \sim 1.94 \times 10^{10}$ | $C_{\text{PAC-CCSD}} \sim 2.14 \times 10^5$ | 106619 |
| PAC* | $2\ C_{\text{QR}} \sim 1.07 \times 10^4$ | $C_{\text{PAC*-CCSD}} \sim 2.51 \times 10^5$ | 1 |

$C_{\text{QR}}$ refers to the FLOPs for the QR decomposition of a complex-valued matrix [16]. Symbol rate processing computational complexities are from the numerical simulation for $n=m=10$ with (8,24,32) star 64 QAM at the SNR of 24 dB (Fig. 4).

that PAC*-CCSD provides a comparable symbol rate detection complexity while its overhead is significantly reduced. As a result, it is found that the proposed PAC*-CCSD always has a net gain to PINV-SESD.

*D. Computational overhead for the modified PAC*

We assume that Givens rotation, a well-known unitary transform based method, is used for QR decompositions. We do not consider the Gram Schmidt method here for it requires costly operations, such as square root operations and divisions, and it is not numerically stable; see [17] and references therein.

There are mainly two kinds of operations in Givens rotation, rotation and cancellation. Rotation makes a complex valued element turned into a real valued one. Cancellation makes a real valued element annihilated. For each columns of $\mathbf{H}'$, rotations and cancellations are performed. For the $k^{th}$ column, $\mathbf{H}'_{\_,k}$, the elements $H'_{i,k}$ for $k \leq i \leq n$ are rotated and become real valued. The complexity for rotations for the column is $(n-k+1) \cdot C_{rot}(n-k+1)$ where $C_{rot}(n-k+1)$ is the computational complexity for a rotation for the $k^{th}$ column. Note that the single rotation for $H'_{i,k}$ amounts to the rotations for the $n-k+1$ elements, $H'_{i,j}$ for $k \leq j \leq n$, since the operations corresponding to the rotation for $H'_{i,k}$ are performed to those elements $H'_{i,j}$ where $k+1 \leq j \leq n$.

Once the rotations are done, the elements $H'_{i,k}$ for $k+1 \leq i \leq n$ which now have turned into real valued ones are annihilated and zero valued. The complexity for cancellations for the column is $(n-k) \cdot C_{can}(n-k+1)$ where $C_{can}(n-k+1)$ is the computational complexity for a cancellation for the $k^{th}$ column. Note also that the single cancellation for $H'_{i,k}$ amounts to the cancellation for the $n-k+1$ elements for the same reason with the rotations. Now, the diagonal element $H'_{k,k}$ becomes real valued and the lower diagonal elements becomes zero. By performing the rotations and the cancellations from the $1^{st}$ to the $n^{th}$ column of $\mathbf{H}'$, a QR decomposition is done. The cost of this operation is
$$C_{QR} = \sum_{k=1}^{n}(n-k+1) \cdot C_{rot}(n-k+1) + (n-k) \cdot C_{can}(n-k+1).$$

There are $n$ QR decompositions required for the proposed variation of PAC. This may be done by performing $n$ separate QR decompositions with a cost of $nC_{QR}$. Fortunately, this can be done very efficiently (with less than $2C_{QR}$) by *i)* performing a QR decomposition on $\mathbf{H}'$ whose columns are ordered by $\delta_i^2$ and *ii)* deriving remaining $n$-1 QR decompositions from the one in step '*i)*'. For the explanation of the method, we assume that $\delta_1^2 \geq \delta_2^2 \geq \cdots \geq \delta_n^2$. We use $\mathbf{H}'_{i_n}$ to denote the reorganized $\mathbf{H}$ by (11). For step '*i)*', in this case, $i_n = n$, a separate QR decomposition is performed. And for the remaining QR decompositions, the upper triangular matrix $\mathbf{R}'_n$ obtained in step '*i)*' is used instead of $\mathbf{H}'$. For the case of $i_n = i$, the $i^{th}$ column of $\mathbf{R}'_n$ is placed at the right-most position, and the columns from the $i+1^{th}$ to the $n^{th}$ positions are shifted to the left by one. Then, the QR decomposition for $i_n = i$ can be done with only $C_i = \sum_{k=i}^{n} C_{rot}(n-k+1) + \sum_{k=i}^{n-1} C_{can}(n-k+1)$. This is done for $i_n = 1, 2, \cdots, n-1$. Then, the net computational overhead for the $n$ QR decompositions becomes $C_{net} = C_{QR} + \sum_{k=1}^{n} k \cdot C_{rot}(n-k+1) + (k-1) \cdot C_{can}(n-k+2) < 2C_{QR}$. [2] An example for the QR decompositions is given in Fig. 3.

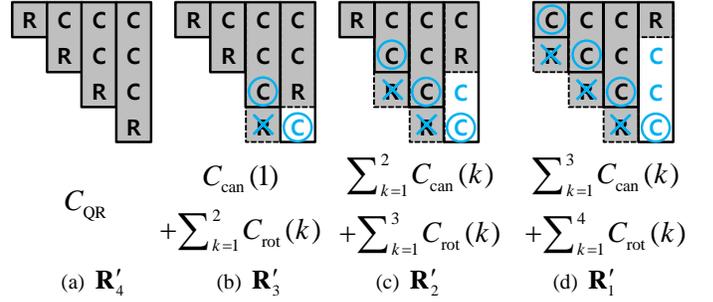

Fig. 3. Efficient QR decompositions for $n$=4. Symbols R and C in the matrix are meant to imply that the element in the pertinent position is either real or complex valued respectively. The circles and crosses represent the rotations and the cancellations, respectively. Gray colored rectangles are the columns taken from the upper triangular matrix in (a).

## VI. EXTENSION TO LIST SPHERE DETECTION

In the iterative detection and decoding (IDD) system, the detector and the decoder exchange soft information repeatedly and improve the quality of their outputs. If messages are encoded with a channel code, and interleaved before they are mapped to modulation symbols $\mathbf{s}$, it is possible to achieve near channel capacity by employing IDD in the MIMO receiver [9], [18], [19].

In this section, we focus on the detection in IDD and consider an application of CSD for it. The channel code and the corresponding decoding operations can be any of those for well performing channel codes, such as low-density parity-check (LDPC) and turbo codes; we do not specify them in this paper. In the detection, a posteriori log-likelihood ratio (AP-LLR) on the all the coded bits that constitute $\mathbf{s}$ are computed for soft information exchange [9]. Note that the coded bits here mean the coded and interleaved bits. The AP-LLR of the $k^{th}$ coded bit, $c_k$, is given by

$$\begin{aligned}L(c_k | \mathbf{r}) &= \ln \frac{P(c_k=1|\mathbf{r})}{P(c_k=0|\mathbf{r})} = \ln \frac{\sum_{\mathbf{s} \in \mathcal{S}_k^1} P(\mathbf{r}|\mathbf{s})P(\mathbf{s})}{\sum_{\mathbf{s} \in \mathcal{S}_k^0} P(\mathbf{r}|\mathbf{s})P(\mathbf{s})} \\ &= L(c_k) + \ln \frac{\sum_{\mathbf{s} \in \mathcal{S}_k^1} P(\mathbf{r}|\mathbf{s}) \prod_{i=1, i \neq k}^{n \log_2 L} P(c_i = c_i(\mathbf{s}))}{\sum_{\mathbf{s} \in \mathcal{S}_k^0} P(\mathbf{r}|\mathbf{s}) \prod_{i=1, i \neq k}^{n \log_2 L} P(c_i = c_i(\mathbf{s}))} \quad (12)\\ &= L(c_k) + \ln \frac{\sum_{\mathbf{s} \in \mathcal{S}_k^1} \varphi(\mathbf{s})}{\sum_{\mathbf{s} \in \mathcal{S}_k^0} \varphi(\mathbf{s})},\end{aligned}$$

---
[2] The computation overhead for $\mathbf{Q}$ matrix generation is not discussed here. However, it is still $C_{net} < 2C_{QR}$ for the computational overhead for $\mathbf{Q}$ matrix generation is proportional to the number of rotations and cancellations.

where $\mathcal{S}_k^1 = \{\mathbf{s} \mid c_k(\mathbf{s}) = 1\}$, $\mathcal{S}_k^0 = \{\mathbf{s} \mid c_k(\mathbf{s}) = 0\}$, $c_i(\mathbf{s})$ is the $i^{\text{th}}$ coded bit of symbol $\mathbf{s}$, $L(c_k) = \ln \frac{P(c_k=1)}{P(c_k=0)}$, and

$$\varphi(\mathbf{s}) = \exp\left(\frac{-1}{2\sigma^2} d(\mathbf{s})\right) \exp\left(\sum_{i=1, i\neq k}^{n\log_2 L} \ln P(c_i = c_i(\mathbf{s}))\right). \quad (13)$$

The direct AP-LLR in (12) without approximations requires the exhaustive Euclidean distance calculations of $d(\mathbf{s})$ for all the candidates $\mathbf{s}$, which are the most complex operations in the calculations. The computation is infeasible for the systems with large $n$ and/or $L$.

SD algorithms which are called the list SD (LSD) [9], [18], [19] address this problem by searching a small set of candidate symbols over which the AP-LLRs are estimated. The approximate AP-LLR is given by

$$L(c_k \mid \mathbf{r}) \approx L(c_k) + \ln\left(\sum_{\mathbf{s} \in \mathcal{L} \cap \mathcal{S}_k^1} \varphi(\mathbf{s}) \Big/ \sum_{\mathbf{s} \in \mathcal{L} \cap \mathcal{S}_k^0} \varphi(\mathbf{s})\right), \quad (14)$$

where $\mathcal{L}$ is the chosen small set of the promising candidates. Hochwald and Brink proposed the $N$-best LSD which chooses candidates $\mathbf{s}$ with $N$ smallest $d(\mathbf{s})$ and stores them into $\mathcal{L}$ [9]. A probabilistic tree pruning method for an approximation of the $N$-best LSD is proposed in [18]. Hochwald and Brink also presented a max-log approximation of AP-LLR for further size reduction on $\mathcal{L}$ [9]. In [19], a modified max-log approximation which considers only the symbols $\mathbf{s}_{\text{ML}}$ and all its binary complements are presented. Note that, in LSDs, small numbers of candidates $\mathbf{s}$ with small $d(\mathbf{s})$ are sought for $\mathcal{L}$. The ground for this is that the significant contribution to the value of $\varphi(\mathbf{s})$ comes from $\mathbf{s}$ with small $d(\mathbf{s})$, and the AP-LLR estimation over the list approximates the direct one. In this sense, given that $|\mathcal{L}| = N$, the $N$-best LSD is the optimal for AP-LLR approximations.

We here show that the use of CSD techniques are beneficial not only to ML search but also to the AP-LLR computation. The AP-LLR computation in (14) is largely made of evaluation of $\varphi(\mathbf{s})$ for $\mathbf{s} \in \mathcal{L}$. The main operations for $\varphi(\mathbf{s})$ in IDD detector is to calculate the $d(\mathbf{s})$ for $\mathbf{s} \in \mathcal{L}$; note that the values for $L(c_k)$ in (12) and $\ln P(c_i = c_i(\mathbf{s}))$ in (13) are given by the decoder. The $d(\mathbf{s})$ for $\mathbf{s} \in \mathcal{L}$ is obtained in the process of LSD, thus, the detection complexity in IDD is mostly from the LSD operations. Therefore, the computational complexity of LSD almost determines the complexity of the IDD detection.

We extend the CSD scheme to LSD and reduce the complexity of $N$-best LSD. We employ the CC for prescreening candidates in LSD. We modify the CSD procedure so that the radius update is performed only when the CSD finds at least $N$ candidates inside the sphere. We found that the proposed list CSD (LCSD) reduces the complexity of the LSD for MIMO systems with general 2D signal constellations without compromising the $N$-best optimality. It will be shown in Section VIII that the considerable complexity reduction is obtained in $N$-best list search using proposed LCSDs. The modification to CSD algorithm for $N$-best list search is provided in TABLE III.

TABLE III
**The $N$-Best List CSD Algorithm: Step 7'**

**Step 7':** (Forward)
    **if** $k = 1$, Insert$(\mathcal{L}, \mathbf{s})$.
        **if** $|\mathcal{L}| = N$, $C \leftarrow d_{\max}^{\mathcal{L}}$ (radius updates), $\hat{\mathcal{L}} \leftarrow \mathcal{L}$, $found \leftarrow 1$.
        go to step 3.
    **else** $k \leftarrow k - 1$, go to step 4
\* Insert$(\mathcal{L}, \mathbf{s})$: insert $\mathbf{s}$ into the set $\mathcal{L}$,
  $d_{\max}^{\mathcal{L}}$: the current maximum $d(\mathbf{s})$ among those of $\mathbf{s} \in \mathcal{L}$.

## VII. COMPLEXITY ANALYSIS

In this section, we aim to analyze the complexity of the proposed CSD in Sec. IV-B. We use FLOPs as a measure of complexity. A lower bound on the expected FLOPs as function of $n$ and SNR is analyzed. On the one hand, a lower bound analysis could be undesirable if the bound is not tight. But, on the other hand, the final expression of the lower bound could be simple enough to give a clear insight. To rip the benefit of the latter, we take the lower bound approach in this paper. We found that the lower bound can still give useful information on the additional complexity reduction behavior of CSD with respect to CV-SD. There are other lower bound analyses in the literature, those by Jaldén and Ottersten [20] and Shim and Kang [14]. Hassibi and Vikalo provide the exact expected complexity [21]. But, their final results are not easy to interpret for their complex expressions as they include integrations.

### A. Overview of CSD complexity

Let the number of nodes which satisfy the SC at the $k^{\text{th}}$ level of the tree be called $N_k^{sc}$, i.e.,

$$N_k^{sc} := \left|\left\{\mathbf{s}_{k:n} \in \mathcal{O}^{n-k+1} \middle| d_k(\mathbf{s}_{k:n}) \leq C\right\}\right|, \quad (15)$$

where $d_k(\mathbf{s}_{k:n}) = \sum_{i=k}^{n} \left|\mathbf{R}_{i,i:n}(\overline{\mathbf{s}}_{i:n} - \mathbf{s}_{i:n}) + \tilde{v}_i\right|^2$ denotes the PED of $\mathbf{s}_{k:n}$ and $\tilde{\mathbf{v}} := \mathbf{Q}^*\mathbf{v}$ is a random vector whose entries are i.i.d. CSCG random variables, and the number of nodes which satisfy the CC at the $k^{\text{th}}$ level of the tree be called $N_k^{cc}$,

$$N_k^{cc} := \left|\left\{s_k \in \mathcal{O} \middle| \tilde{\Delta}_k(s_k) \leq C\right\}\right|, \quad (16)$$

where $\tilde{\Delta}_k(s_k) := \delta_k^{-2} |x_k - s_k|^2 = \delta_k^{-2} \left|\overline{s}_k - s_k + \mathbf{R}_{k,\_}^{\dagger} \tilde{\mathbf{v}}\right|^2$ denotes the C-metric of $s_k$ divided by $\delta_k^2$.

In CV-SD, the complexity is given by

$$\mathbb{C}_{\text{SD-}\mathbb{C}} = \sum_{k=1}^{n} \left(6(n-k) + 9L\right) N_{k+1}^{sc}, \quad (17)$$

where $N_{n+1}^{sc} = 1$. Recall the $L$-expansion property (Sec. III-B). That is, for the identification of $N_k^{sc}$ nodes at the $k^{\text{th}}$ level of the tree, $L \cdot N_{k+1}^{sc}$ PED computations are required; this results in high complexity in the CV-SD; note the $L$ in $9L$.

In CSD, the high complexity problem of the CV-SD is alleviated. Among all the possible children nodes of each surviving node at the previous level, $k+1$, note that there are $L$ children nodes, only those children nodes, $N_k^{cc} \leq L$ of them, that satisfy CC are passed for PED computations. The complexity for CSD is given by

$$\mathcal{C}_{\text{CSD-C}} = \sum_{k=1}^{n} 6L + \left(6(n-k) + 9N_k^{cc}\right) N_{k+1}^{sc}. \quad (18)$$

Here, $6L$ is the FLOPs for the CC tests at each level.

We analyze the complexity of CSD in the following subsection.

### B. CSD complexity

The expected complexity of CSD over $\mathbf{H}$, $\overline{\mathbf{s}}$, and $\mathbf{v}$ is

$$\mathrm{E}[\mathcal{C}_{\text{CSD-C}}] = \sum_{k=1}^{n} 6L + 6(n-k) \cdot \mathrm{E}\left[N_{k+1}^{sc}\right] + 9 \cdot \mathrm{E}\left[N_k^{cc} N_{k+1}^{sc}\right]. \quad (19)$$

For simplicity of the derivation, we assume $m=n$ in the sequel. But, it can be easily seen that the result for the general case $m>n$ remains the same. We also use the SNR $\rho = \frac{m}{\sigma^2}$ and the radius $C = \alpha n \sigma^2$ where $\alpha$ is determined so that a solution is found with a high probability, $1-\varepsilon = 0.99$, in the sphere [21]. Since the complexity is averaged over $\overline{\mathbf{s}}$, we consider $\tilde{\Delta}_k$ and $d_k$ as functions of not only $\mathbf{s}$ but also $\overline{\mathbf{s}}$.

*Lemma 4:* $\mathrm{E}[N_k^{cc} N_{k+1}^{sc}]$ is lower bounded by

$$\mathrm{E}[N_k^{cc} N_{k+1}^{sc}] \geq \frac{1}{L^{n-k+1}} \sum_{\mathbf{s}_{k:n}} \sum_{\overline{\mathbf{s}}_{k:n}} \Pr\left(\tilde{\Delta}_k(s_k, \overline{s}_k) \leq C\right) \Pr\left(d_{k+1}(\mathbf{s}_{k+1:n}, \overline{\mathbf{s}}_{k+1:n}) \leq C\right). \quad (20)$$

*Proof:* See Appendix A. □

We compute $\Pr\left(\tilde{\Delta}_k(s_k, \overline{s}_k) \leq C\right)$ first, and then $\Pr\left(d_{k+1}(\mathbf{s}_{k+1:n}, \overline{\mathbf{s}}_{k+1:n}) \leq C\right)$.

*Lemma 5:* The probability $\Pr\left(\tilde{\Delta}_k(s_k, \overline{s}_k) \leq C\right)$ is lower bounded by

$$\Pr\left(\tilde{\Delta}_k(s_k, \overline{s}_k) \leq C\right) \geq \left(1 - \frac{1}{\alpha n}\left(\left|\overline{s}_k - s_k\right|^2 \rho + 1\right)\right)_+. \quad (21)$$

*Proof:* See Appendix B. □

Now, we compute a lower bound on $\Pr\left(d_{k+1}(\mathbf{s}_{k+1:n}, \overline{\mathbf{s}}_{k+1:n}) \leq C\right)$.

*Lemma 6:* The probability $\Pr\left(d_{k+1}(\mathbf{s}_{k+1:n}, \overline{\mathbf{s}}_{k+1:n}) \leq C\right)$ is lower bounded by

$$\Pr\left(d_{k+1}(\mathbf{s}_{k+1:n}, \overline{\mathbf{s}}_{k+1:n}) \leq C\right) \geq \left(1 - \frac{n-k}{\alpha n}\left(\left\|\overline{\mathbf{s}}_{k+1:n} - \mathbf{s}_{k+1:n}\right\|^2 \frac{\rho}{n} + 1\right)\right)_+. \quad (22)$$

*Proof:* See Appendix C. □

Finally, we obtain a lower bound on the expected complexity of CSD.

*Theorem 7:* The expected complexity $\mathrm{E}[\mathcal{C}_{\text{CSD-C}}]$ is lower bounded by

$$\mathrm{E}[\mathcal{C}_{\text{CSD-C}}] \geq \sum_{k=1}^{n} 6L + 6(n-k)(1-\beta_s)_+ + 9L^{n-k+1}(1-\beta_s)_+(1-\beta_c)_+, \quad (23)$$

where

$$\beta_c := \frac{1}{\alpha n}\left(\mathrm{E}\left[\left|\overline{s}_k - s_k\right|^2\right]\rho + 1\right) \quad (24)$$

and $\beta_s := \frac{n-k}{\alpha n}\left(\mathrm{E}\left[\left|\overline{s}_k - s_k\right|^2\right]\frac{n-k}{n}\rho + 1\right)$ are the complexity reduction factors which are determined by the system parameters. Here, $\mathrm{E}\left[\left|\overline{s}_i - s_i\right|^2\right]$, the average intra-constellation squared distance, is determined when the constellation is decided; for example, it is 2 for QAMs and PSKs, and 1.8163 for (8,24,32) 64 star QAM.

*Proof:* See Appendix D. □

A lower bound on the expected complexity of the CV-SD can be easily obtained by using a similar procedure, but with (17) and Lemma 6.

*Theorem 8:* The expected complexity $\mathrm{E}[\mathcal{C}_{\text{SD-C}}]$ is lower bounded by

$$\mathrm{E}[\mathcal{C}_{\text{SD-C}}] \geq \sum_{k=1}^{n} 6(n-k)(1-\beta_s)_+ + 9L^{n-k+1}(1-\beta_s)_+. \quad (25)$$

Here, we see that the additional constraint CC in CSD results in an additional multiplicative factor $(1-\beta_c)_+$, in the last term of the CSD complexity in (23), compared to the CV-SD complexity in (25). Among the last term in (25), $\beta_c$ portion of them are excluded in (23); $(\,)_+$ and $6L$ are ignored for easy evaluation. We expect that this gives CSD a complexity reduction relative to the CV-SD.

We note from (24) that the additional complexity reduction factor $\beta_c$ increases as *i)* SNR increases and/or *ii)* $n$ decreases. That is, it is expected that the additional complexity reduction in CSD to the CV-SD become more as SNR increases and/or $n$ decreases. This finding exactly matches to the complexity reduction behavior of CSD which is observed in the simulations. Here, the complexity reduction behavior with respect to $n$ may not be attractive. But, this problem disappears when PAC is utilized; it performs better as $n$ increases (Sec. VIII).

## VIII. SIMULATION RESULTS

### A. Setup

In this section, we show the complexity reduction capability of the proposed CSDs through system simulations. We compare the proposed CSDs (CSD in Sec. IV-B, C-CSD in Sec. IV-C, and PAC-CCSD in Sec. V) with the conventional CV-SDs (SD, SE-SD, and PINV-SESD) which are also directly applicable to general complex valued constellations. We also compare the complexities of the proposed LCSDs in Sec. VI to those of the conventional LSDs for $N$-best list search. We considered star QAM, rectangular QAM, and PSK in simulations. However, we present the result for star QAM only. The patterns of the complexity reduction of the proposed CSDs in rectangular QAM and PSK are almost the same to those in star QAM. Note that the usage of CSD is not limited to these three constellations, and is applicable to any arbitrary complex valued constellation. We use $8\times 8$ and $10\times 10$ MIMO systems with (4, 24, 32) star 64 QAM for the ring ratios of (2, 3) [5].[3] The initial radius is set to $C_0 = \alpha n \sigma^2$. We employ the average FLOPs as a metric for complexity; they are averaged over $10^4$ runs of channels in each SNR value. The SNR range is determined by considering the dynamic range of the FLOPs so that the FLOPs at the minimum considered SNR and those at the maximum SNR are reasonably far from each other. If they are too apart, the FLOPs at the maximum SNR all look overlapped and not distinguishable.

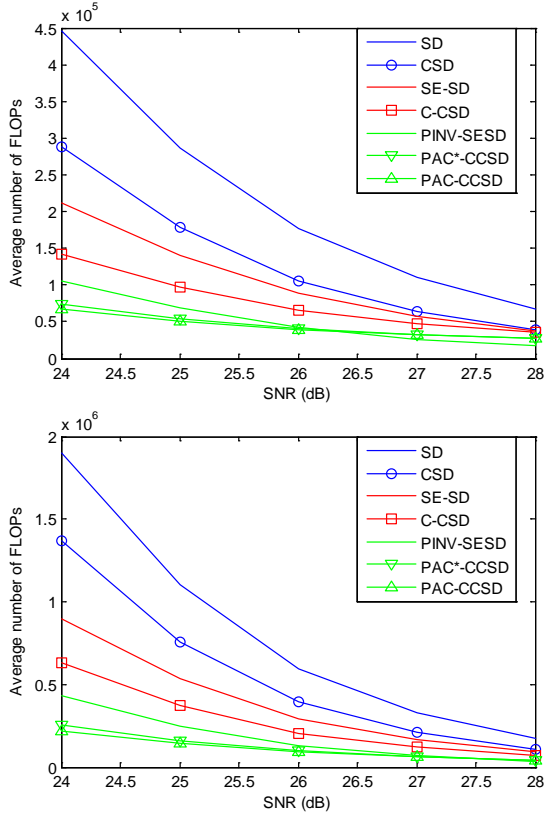

Fig. 4. Complexities of CV-SDs and the proposed CSDs for 8×8 and 10×10 MIMO systems with (8,24,32) star 64 QAM.

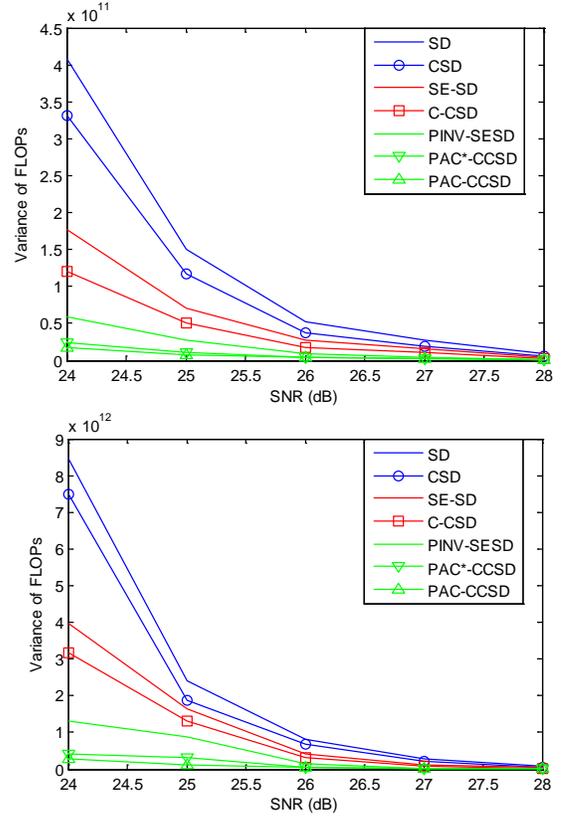

Fig. 5. Complexity variances of CV-SDs and the proposed CSDs for 8×8 and 10×10 MIMO systems with (8,24,32) star 64 QAM.

### B. Results

Fig. 4 plots the average numbers of FLOPs for the CV-SDs and the proposed CSDs in MIMO systems with (8,24,32) star 64 QAM. The proposed CSDs reduce the complexities of the CV-SDs considerably. We observe that CSD, C-CSD, and PAC-CCSD outperform SD, SE-SD, and PINV-SESD by 35%, 37%, 41%, 42%, 43%, by 33%, 30%, 27%, 17%, 3%, and by 37%, 26%, 6%, -18%, -37%,[4] respectively, at SNR (dB) of 24, 25, 26, 27, 28 in 8×8 MIMO systems. For 10×10 MIMO systems, CSD, C-CSD, and PAC-CCSD reduce the complexities of SD, SE-SD, and PINV-SESD by 28%, 31%, 33%, 36%, 40%, by 30%, 31%, 29%, 27%, 21%, and by 50%, 43%, 30%, 13%, -14%, respectively.

---
[3] The ring ratios indicate the ratios of the minimum ring amplitude to the other ring amplitude.
[4] The minus sign means that the complexity of the proposed method is more than that of the conventional method. The corresponding numbers are the increased amounts of complexity compared to those of the conventional method.

As discussed in Section VII-B, the complexity reduction factor of CSD increases as SNR increases and/or $n$ decreases. Interestingly, the complexity reduction factor of PAC-CCSD behaves in an opposite way to that of the CSD; it increases as SNR decreases and $n$ increases. There are significant complexity reductions in the proposed CSDs in almost all the SNR regions (up to 43% by CSD at 28 dB and up to 37% by PAC-CCSD at 24 dB in 8×8 MIMO systems and up to 40% by CSD at 28 dB and up to 50% by PAC-CCSD at 24 dB in 10×10 MIMO systems). Note that these complexity reductions in CSDs are obtained without compromising its ML performance; of course, they all achieved the ML performance in simulations. It was found that PAC-CCSD has a higher complexity than PINV-SESD does at SNRs of 27 dB and 28 dB in a 8×8 MIMO system and at SNR of 28 dB in a 10×10 MIMO system. Still, it is not a big issue since both exhibit low complexities in this SNR region.

The proposed CSDs also reduce the complexity variances of the CV-SDs (Fig. 5). Note that in Fig. 5 variance reductions in the proposed CSDs are significant throughout the entire SNR range considered. For example, the PAC-CCSD scheme

Submitted to IEEE trans. on Vehicular Technology

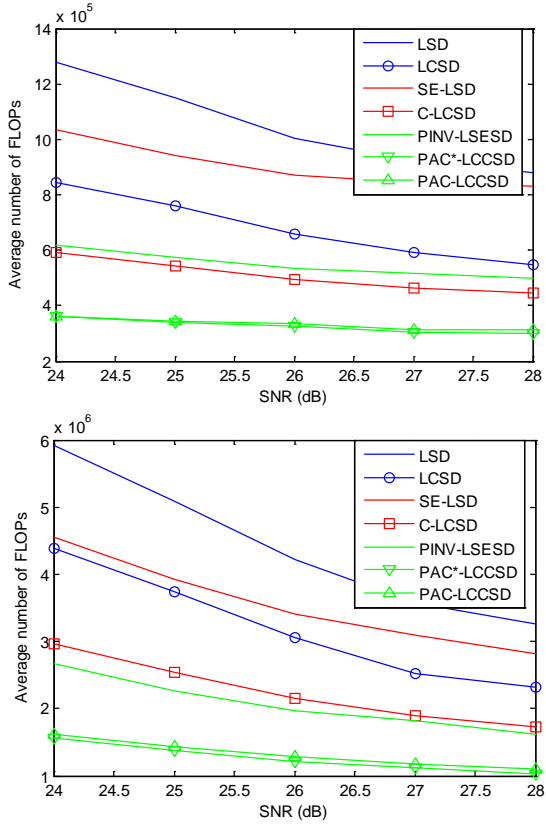

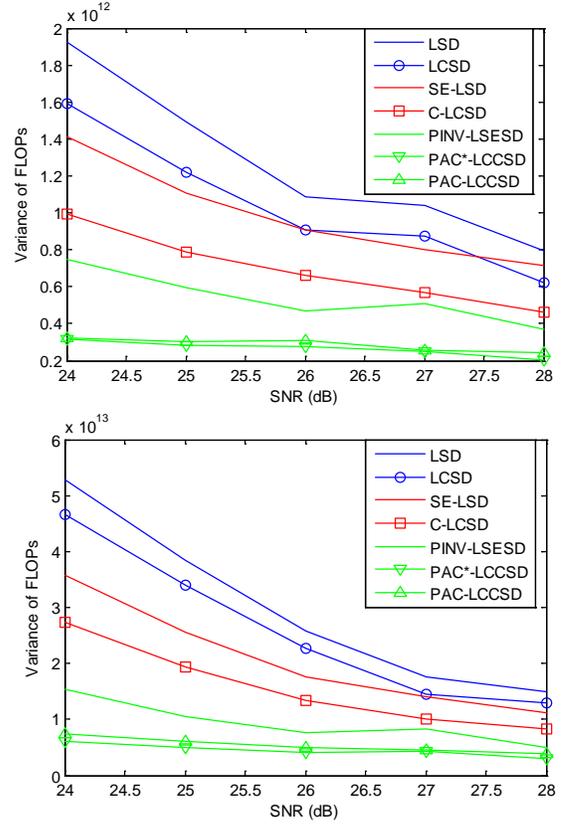

Fig. 6. Complexities of CV-LSDs and the proposed LCSDs for 8×8 and 10×10 MIMO systems with (8,24,32) star 64 QAM.

Fig. 7. Complexity variances of CV-LSDs and the proposed LCSDs for 8×8 and 10×10 MIMO systems with (8,24,32) star 64 QAM.

provides reductions up to 71% in a 8×8 MIMO system and up to 87% in a 10×10 MIMO system.

We also consider $N$-best list detection. The patterns of the complexity reduction of the proposed LCSDs are similar to those of CSDs but with more margins. It is observed that the proposed LCSDs perform better than LSDs by large margins. For 64 QAM, LCSD, C-LCSD, and PAC-LCCSD outperform LSD, SE-LSD, and PINV-LSESD in terms of average number of FLOPs by 34%, 34%, 34%, 36%, 37%, by 43%, 43%, 43%, 45%, 47%, and by 41%, 40%, 37%, 39%, 37%, respectively, in a 8×8 system with $N$=20, by 26%, 27%, 27%, 29%, 29%, by 35%, 35%, 37%, 39%, 39%, and by 39%, 37%, 35%, 36%, 32%, respectively, in 10×10 system with $N$=30 at SNR (dB) of 24, 25, 26, 27, 28 (Fig. 6). In terms of complexity variance, it is shown in Fig. 7 that the variance reductions by LCSDs are significant (up to 57% in a 8×8 MIMO system and up to 51% in a 10×10 MIMO system) as they are in CSDs.

## IX. CONCLUSION

In this paper, we have proposed CSD, a low complexity CV-SD, for general 2D constellations. CSD uses the simple circular constraint (CC) for prescreening candidates and pruning some of them even before executing the SC tests. Simulations have shown that CSD yields a large reduction in the number of FLOPs. We also propose a further complexity reduction strategy, the Predict-And-Change (PAC). PAC also provides a further considerable complexity reduction. Thanks to the proposed methods, the CSD becomes surely a good candidate for a general 2D constellation low complexity MIMO detector. It was also shown that the proposed methods are beneficial in soft output SD schemes. With the proposed CSD, it becomes possible to decode signals with any integer data rate (not only for $L = 2^i$, $i = 1, 2, \cdots$) and arbitrary shape of 2D constellations which may be optimal for target applications with low complexity while achieving the optimal error rate performance as CSD is compatible to general 2D constellation. A large number of 2D constellations can be handled in this single CSD algorithm without any additional constellation-dependent functionality.

## APPENDIX

### A. Proof of Lemma 4

$$\begin{aligned}
& \mathrm{E}[N_k^{cc} N_{k+1}^{sc}] \\
&= \mathrm{E}\left[\sum_{s_k} I\{\tilde{\Delta}_k(s_k, \overline{s}_k) \leq C\} \sum_{\mathbf{s}_{k+1:n}} I\{d_{k+1}(\mathbf{s}_{k+1:n}, \overline{\mathbf{s}}_{k+1:n}) \leq C\}\right] \\
&= \sum_{\mathbf{s}_{k:n}} \Pr\left(\tilde{\Delta}_k(s_k, \overline{s}_k) \leq C, d_{k+1}(\mathbf{s}_{k+1:n}, \overline{\mathbf{s}}_{k+1:n}) \leq C\right) \\
&\stackrel{(a)}{=} \frac{1}{L^{n-k+1}} \sum_{\mathbf{s}_{k:n}} \sum_{\overline{\mathbf{s}}_{k:n}} \Pr\left(\tilde{\Delta}_k(s_k, \overline{s}_k) \leq C, d_{k+1}(\mathbf{s}_{k+1:n}, \overline{\mathbf{s}}_{k+1:n}) \leq C\right),
\end{aligned} \quad (26)$$

where $I\{\cdot\}$ is the indicator function which is 1 for the condition inside the bracket is true, otherwise 0, and (a) is from $\Pr(\bar{\mathbf{s}}_{k:n}) = 1/L^{n-k+1}$.

Before proceeding further, we introduce a definition and a lemma which are useful for further simplification of (26).

*Definition 9:* (*non-negative correlation*) Two random events $\mathcal{A}$ and $\mathcal{B}$ are said to be *non-negatively correlated* if $\mathrm{cov}(I\{\mathcal{A}\}, I\{\mathcal{B}\}) \geq 0$. We use the tilde symbol ~ between two events, *i.e.*, $\mathcal{A} \sim \mathcal{B}$, to imply that the two are *non-negatively correlated*.

*Lemma 10:* $\mathcal{A} \sim \mathcal{B}$ if and only if $\Pr(\mathcal{AB}) \geq \Pr(\mathcal{A})\Pr(\mathcal{B})$.

*Proof:* This can be easily seen by Def. 9. □

Here we aim to show that $I\{d_{k+1}(\mathbf{s}_{k+1:n}, \bar{\mathbf{s}}_{k+1:n}) \leq C\}$ $\sim I\{\tilde{\Delta}(s_k, \bar{s}_k) \leq C\}$. To this end, we first show that $I\{d_{k+1} \leq C\} \sim I\{d_1 \leq C\}$, and then $I\{d_1 \leq C\} \sim I\{\tilde{\Delta}_k \leq C\}$, and thus $I\{d_{k+1} \leq C\} \sim I\{\tilde{\Delta}_k \leq C\}$. Before we move on, note the inequalities $d_{k+1} \leq d_1$ and $\tilde{\Delta}_k \leq d_1$ from (3) and (5). Now, we show that $I\{d_{k+1} \leq C\} \sim I\{d_1 \leq C\}$ by showing the following,

$$\Pr(d_{k+1} \leq C, d_1 \leq C) = \Pr(d_1 \leq C) \geq \Pr(d_{k+1} \leq C)\Pr(d_1 \leq C).$$

Thus, the first is shown. Now, for the second, we show that $\Pr(d_1 \leq C, \tilde{\Delta}_k \leq C) = \Pr(d_1 \leq C) \geq \Pr(d_1 \leq C)\Pr(\tilde{\Delta}_k \leq C)$, thus $I\{d_1 \leq C\} \sim I\{\tilde{\Delta}_k \leq C\}$. Therefore, $I\{d_{k+1}(\mathbf{s}_{k+1:n}, \bar{\mathbf{s}}_{k+1:n}) \leq C\} \sim I\{\tilde{\Delta}_k(s_k, \bar{s}_k) \leq C\}$; this is also verified through extensive simulations.

Now, return to the discussion of (26). Let $\mathcal{A} := \{\tilde{\Delta}_k(s_k, \bar{s}_k) \leq C\}$ and $\mathcal{B} := \{d_{k+1}(\mathbf{s}_{k+1:n}, \bar{\mathbf{s}}_{k+1:n}) \leq C\}$. Using Lemma 10, the joint probability $\Pr(\tilde{\Delta}_k(s_k, \bar{s}_k) \leq C, d_{k+1}(\mathbf{s}_{k+1:n}, \bar{\mathbf{s}}_{k+1:n}) \leq C)$ is lower bounded by $\Pr(\tilde{\Delta}_k(s_k, \bar{s}_k) \leq C)\Pr(d_{k+1}(\mathbf{s}_{k+1:n}, \bar{\mathbf{s}}_{k+1:n}) \leq C)$.

### B. Proof of Lemma 5

$\mathrm{E}\left[\tilde{\Delta}_k(s_k, \bar{s}_k)\right]$ is expressed as follows

$$\mathrm{E}\left[\frac{|\bar{s}_k - s_k + \mathbf{R}^{-1}_{k,\_}\tilde{\mathbf{v}}|^2}{\|\mathbf{R}^{-1}_{k,\_}\|^2}\right]$$

$$= \frac{|\bar{s}_k - s_k|^2}{\mathrm{E}\left[\|\mathbf{R}^{-1}_{k,\_}\|^2\right]} + \mathrm{E}\left[\frac{\left|\sum_{i=1}^{n} R^{-1}_{k,j} \cdot \tilde{v}_i\right|^2}{\|\mathbf{R}^{-1}_{k,\_}\|^2}\right] + 2(\bar{s}_k - s_k)^* \mathrm{E}\left[\frac{\sum_{i=1}^{n} R^{-1}_{k,j} \cdot \tilde{v}_i}{\|\mathbf{R}^{-1}_{k,\_}\|^2}\right].$$

The first term is upper bounded as follows,

$$|\bar{s}_k - s_k|^2 / \mathrm{E}\left[\|\mathbf{R}^{-1}_{k,\_}\|^2\right] \underset{(a)}{\leq} |\bar{s}_k - s_k|^2 \mathrm{E}\left[\|\mathbf{R}_{\_,k}\|^2\right] \underset{(b)}{=} |\bar{s}_k - s_k|^2 \cdot n,$$

where (a) is from $\mathrm{E}\|\mathbf{R}^{-1}_{k,\_}\|^2 \mathrm{E}\|\mathbf{R}_{\_,k}\|^2 \geq \mathrm{E}|\mathbf{R}^{-1}_{k,\_}\mathbf{R}_{\_,k}| = 1$ and (b) is from $\|\mathbf{R}_{\_,k}\|^2 = \|\mathbf{Q}^*\mathbf{H}_{\_,k}\|^2 = \|\mathbf{H}_{\_,k}\|^2$.

The second term becomes $\sigma^2$ as follows,

$$\mathrm{E}\left[\left|\sum_{i=1}^{n} R^{-1}_{k,j} \cdot \tilde{v}_i\right|^2 \Big/ \|\mathbf{R}^{-1}_{k,\_}\|^2\right]$$

$$\underset{(c)}{=} \sum_{i=1}^{n} \mathrm{E}\left[\frac{|R^{-1}_{k,i}|^2}{\|\mathbf{R}^{-1}_{k,\_}\|^2}\right] \mathrm{E}\left[|\tilde{v}_i|^2\right] + \sum_{i=1, j=1, i\neq j}^{n} \mathrm{E}\left[\frac{R^{-1*}_{k,i} R^{-1}_{k,j}}{\|\mathbf{R}^{-1}_{k,\_}\|^2}\right] \mathrm{E}[\tilde{v}_i^*]\mathrm{E}[\tilde{v}_j]$$

$$= \sigma^2,$$

where (c) is from the independence of $\tilde{\mathbf{v}}$ and $\mathbf{R}^{-1}$.

The third term becomes zero as follows

$$\mathrm{E}\left[\sum_{i=1}^{n} R^{-1}_{k,j} \cdot \tilde{v}_i \Big/ \|\mathbf{R}^{-1}_{k,\_}\|^2\right] \underset{(d)}{=} \sum_{i=1}^{n} \mathrm{E}\left[R^{-1}_{k,i} \Big/ \|\mathbf{R}^{-1}_{k,\_}\|^2\right] \mathrm{E}[\tilde{v}_i] = 0,$$

where (d) also comes from the independence of $\tilde{\mathbf{v}}$ and $\mathbf{R}^{-1}$.

Therefore, $\mathrm{E}[\tilde{\Delta}_k(s_k, \bar{s}_k)] \leq \sigma^2\left(|\bar{s}_k - s_k|^2 \rho + 1\right)$.

Now, $\Pr\left(\tilde{\Delta}_k(s_k, \bar{s}_k) \leq C\right)$ is lower bounded by as follows,

$$\Pr\left(\tilde{\Delta}_k(s_k, \bar{s}_k) \leq C\right) \underset{(a)}{\geq} \left(1 - \frac{1}{\alpha n}\left(|\bar{s}_k - s_k|^2 \rho + 1\right)\right)_+,$$

where (a) is from the Markov inequality and $(\cdot)_+$ is to make sure the probability to be nonnegative.

### C. Proof of Lemma 6

$\mathrm{E}\left[d_{k+1}(\mathbf{s}_{k+1:n}, \bar{\mathbf{s}}_{k+1:n})\right]$ is expressed as follows

$$\mathrm{E}\left[d_{k+1}(\mathbf{s}_{k+1:n}, \bar{\mathbf{s}}_{k+1:n})\right]$$

$$= \sum_{i=k+1}^{n} \mathrm{E}\left[\left|\sum_{j=i}^{n} R_{i,j}(\bar{s}_j - s_j) + \tilde{v}_i\right|^2\right]$$

$$\underset{(a)}{=} \sum_{i=k+1}^{n}\sum_{j=i}^{n} |\bar{s}_j - s_j|^2 \left(\mathrm{E}\left[\|\mathbf{R}_{\_,j}\|^2\right] - \sum_{i=1}^{k}\mathrm{E}\left[|R_{i,j}|^2\right]\right) + (n-k)\sigma^2$$

$$\underset{(b)}{=} (n-k)\|\bar{\mathbf{s}}_{k+1:n} - \mathbf{s}_{k+1:n}\|^2 + (n-k)\sigma^2,$$

where (a) comes from the independence between $\tilde{\mathbf{v}}$ and $\mathbf{R}$, and the fact that any off-diagonal element of $\mathbf{R}$ has mean zero [21] and (b) comes from $\|\mathbf{R}_{\_,j}\|^2 = \|\mathbf{H}_{\_,j}\|^2$.

Finally, using the Markov inequality,

$$\Pr\left(d_{k+1}(\mathbf{s}_{k+1:n}, \bar{\mathbf{s}}_{k+1:n}) \leq C\right) \geq \left(1 - \frac{n-k}{\alpha n}\left(\|\bar{\mathbf{s}}_{k+1:n} - \mathbf{s}_{k+1:n}\|^2 \frac{\rho}{n} + 1\right)\right)_+.$$

### D. Proof of Theorem 7

$$\mathrm{E}[N_k^{cc} N_{k+1}^{sc}] \underset{(a)}{\geq} (1/L^{n-k+1})\sum_{s_k}\sum_{\bar{s}_k}\left(1 - \frac{1}{\alpha n}\left(|\bar{s}_k - s_k|^2 \rho + 1\right)\right)_+$$

$$\cdot \sum_{\mathbf{s}_{k+1:n}}\sum_{\bar{\mathbf{s}}_{k+1:n}}\left(1 - \frac{n-k}{\alpha n}\left(\|\bar{\mathbf{s}}_{k+1:n} - \mathbf{s}_{k+1:n}\|^2 \frac{\rho}{n} + 1\right)\right)_+$$

$$\overset{(b)}{=} L^{n-k+1} \cdot \left(1 - \frac{1}{\alpha n}\left(\mathrm{E}\left[\left|\overline{s}_k - s_k\right|^2\right]\rho + 1\right)\right)_+$$

$$\cdot \left(1 - \frac{n-k}{\alpha n}\left(\mathrm{E}\left[\left|\overline{s}_k - s_k\right|^2\right]\frac{n-k}{n}\rho + 1\right)\right)_+,$$

where (a) is from Lemma 4, 5 and 6, (b) is from $\left(\sum_k a_k\right)_+ \leq \sum_k (a_k)_+$, $\sum_{\overline{\mathbf{s}}_{k+1:n}} \sum_{\mathbf{s}_{k+1:n}} 1 = L^{2(n-k)}$, and

$$\mathrm{E}\left[\left\|\overline{\mathbf{s}}_{k+1:n} - \mathbf{s}_{k+1:n}\right\|^2\right] = \frac{1}{L^{2(n-k)}} \sum_{\overline{\mathbf{s}}_{k+1:n}} \sum_{\mathbf{s}_{k+1:n}} \left\|\overline{\mathbf{s}}_{k+1:n} - \mathbf{s}_{k+1:n}\right\|^2.$$

$\mathrm{E}[N_{k+1}^{sc}]$ can be easily derived using a similar procedure.